\documentclass[12pt]{article}
\usepackage{graphicx}
\usepackage{amssymb}
\usepackage{amsmath}
\newsymbol \blackbox 1004
\newcommand{\eh}{\hfill}\newlength{\sperr}

\newenvironment{proof}{{\settowidth{\sperr}{\bf\rm
Proof}%
\par\addvspace{0.3cm}\noindent\parbox[t]{1.3\sperr}
{\bf\rm P\eh r\eh o\eh o\eh f\eh }%
}}{\nopagebreak\mbox{}
$\blackbox$\par\addvspace{0.3cm}}

\def\nn{\nonumber}

\newtheorem{Pa}{Paper}[section]
\newtheorem{Tm}[Pa]{{\bf Theorem}}

\newtheorem{Rk}[Pa]{{\bf Remark}}

\newtheorem{Pn}[Pa]{{\bf Proposition}}
\title{Non-extensive statistical mechanics: Gibbs-type formula,
existence and uniqueness of its solution}
\author{Lev Sakhnovich}
\date{}
\begin{document}
\maketitle
\emph{199 Cove ave., Milford, CT, 06461, USA}
 
 E-mail: lsakhnovich@gmail.com
 
 \begin{abstract} 
 Existence and uniqueness results for the solution
 of the Gibbs-type formula from non-extensive mechanics are derived
 rigorously.  A new conditional extremal problem is proposed to 
 get  in a more simple way the Gibbs-type formula itself.
 
 \end{abstract}
 
 

 \section{Introduction}
 Non-extensive statistical mechanics is  an actively studied domain
 and a spacious bibliography exists already (see some references
 in \cite{GeTs, T, Ts}).
 Following C. Tsallis \cite{T} we define 
 entropy by a basic formula from  non-extensive mechanics:
 \begin{equation}
\label{1} 
 S_{q}=(1-\sum_{i=1}^{n}p_{i}^{q})/(q-1),\quad \sum_{i=1}^{n}p_{i}=1,\quad
 p_{i}>0,\quad q>0,\end{equation}
 where  n is the total number of probabilities.   Energy is defined by the formula (see\cite{T}):
 \begin{equation}
\label{2} 
 U_{q}=(\sum_{i=1}^{n}p_{i}^{q}E_{i})/(\sum_{i=1}^{n}p_{i}^{q}),
 \end{equation}where $E_{i}$ are the eigenvalues of the Hamiltonian of the corresponding system.
 In our approach we consider a new extremal problem. Namely, we fix the Lagrange multiplier
$\beta=1/kT$, that is, we fix the temperature and introduce the
Helmholtz free energy (up to a constant multiple) as the compromise function
$$F(\beta,p_{1},p_{2},...,p_{n})=-{\beta}U_{q}+S_{q}.$$ 
Similar to \cite{LAS, LAS2}
we assume the \\
{\bf Fundamental principle:}
{\it The probabilities $\{p_i\}$ are the coordinates
of the stationary point of the compromise function.}

Hence, the probabilities
$\{p_i\}$ and, correspondingly, the mean energy $U_{q}$ and the entropy $S_{q}$ are obtained as the solution of this extremal problem. 

The suggested approach affords a simple and rigorous treatment of the basic Gibbs-type formula,
being a development of the
papers \cite{LAS} and \cite{LAS2}, where the same approach was applied
to ordinary quantum and classical mechanics, respectively.
Namely, we recall that the stationary point $\tilde{P}=(\tilde{p_{1}},\tilde{p_{2}},...,\tilde{p_{n}})$
of the function $F(\beta,p_{1},p_{2},...,p_{n})$ is a solution of the system
\begin{equation}\label{3}
\frac{\partial{F}(\beta,p_{1},p_{2},...,p_{n})}{\partial{p_{i}}}=0,
\quad 1{\leq}i{\leq}n.\end{equation}
In view of \eqref{3}, it is easy to see  that the fundamental principle implies
equalities
\begin{equation}
\label{4}
\tilde{p_{i}}=\hat{Z}_{q}^{-1}\Big(1+(q-1){\beta}(E_{i}-U_{q})/(\sum_{i=1}^{n}{\tilde
p_{i}}^{q})\Big)^{\frac{1}{1-q}}
,\end{equation}
where
\begin{equation} \label{5}
\hat{Z}_{q}=\sum_{i=1}^{n}\Big(1+(q-1){\beta}(E_{i}-U_{q})/(\sum_{i=1}^{n}{\tilde
p_{i}}^{q})
\Big)^{\frac{1}{1-q}}.
\end{equation}
Thus, the fundamental principle conforms to the basic Gibbs-type
relations from \cite[p. 12]{T}.  There is also an interesting intersection with the results from game theory, which is discussed in the Conclusion.

We note that (differently from the classical Gibbs formula) the Gibbs-type formula \eqref{4}
is, in fact, an equation. Thus, an important problem of the existence and uniqueness of its solution
appear. Further we denote probabilities
satisfying fundamental principle and so satisfying \eqref{4} by
${\tilde{p_i}}$. The next section is dedicated to the rigorous proof of the 
existence and uniqueness of  the solution of \eqref{4}.

\section{Extremum  points}
We introduce the following values
\begin{equation}
\label{6}
E_{\max}=\max\{E_{1},E_{2},...,E_{n}\},\quad
E_{\min}=\min\{E_{1},E_{2},...,E_{n}\}.
\end{equation}
We need  such a solution $\tilde{p}_{i}$ of system (\ref{4}), (\ref{5}), that
\begin{equation}\label{7}
\tilde{p}_{i}>0,\quad 1{\leq}i{\leq}n.\end{equation}
\begin{Pn} \label{Pn1} Let the following conditions
\begin{equation}\label{8}
q>1,\quad
\beta>0,\quad 1-{\beta}(q-1)(E_{\max}-E_{\min})n^{q-1}>0
\end{equation} 
hold. 
Then every solution of system \eqref{4}, \eqref{5} satisfies conditions \eqref{7}.
\end{Pn}

\begin{proof}. The assertion of the proposition follows directly from (1.4),
(1.5) and the inequality
\begin{equation}\nn
z_{q}=\sum_{i=1}^{n}p_{i}^{q}{\geq}n^{1-q}.\end{equation}
\end{proof}
\begin{Pn} \label{Pn2} Let the following conditions
\begin{equation}\label{10}
0<q<1,\quad \beta>0,\quad 1+{\beta}(1-q)(E_{\min}-E_{\max})>0
\end{equation}
hold. Then every solution of system \eqref{4}, \eqref{5} satisfies conditions \eqref{7}.
\end{Pn}
\begin{proof}. The assertion of the proposition follows directly from \eqref{4},
\eqref{5} and the inequality
\begin{equation} \nn
z_{q}=\sum_{i=1}^{n}p_{i}^{q}{\geq}1.
\end{equation}
\end{proof}
\begin{Rk} Proposition  \ref{Pn2} is true in the case $n=\infty$ as well.
\end{Rk}
Let us denote by $D=\{(p_{1},p_{2},...,p_{n})\}$ the set of points, where
\begin{equation}
p_{i}{\geq}0,\quad
1{\leq}i{\leq}n;\quad
\sum_{i=1}^{n}p_{i}=1.\end{equation}
The set $D$ is compact and convex. A topological space X is said to have the fixed point property (briefly FPP) if for any continuous function $f:X{\to}X$
there exists $x{\in}X$  such that f(x) = x.
According to the Brouwer fixed point theorem, every compact and convex subset of a euclidean space has the FPP. 

It is easy to see, that the following statement is true.
\begin{Pn} \label{Pn3}
Let the conditions of either Proposition \ref{Pn1}
or Proposition \ref{Pn2} hold. Then the relations
\begin{equation}\label{j1}
r_{i}=\hat{Z}_{q}^{-1}\Big(1+(q-1){\beta}(E_{i}-U_{q})/(\sum_{i=1}^{n}p_{i}^{q})\Big)^{\frac{1}{1-q}}
,\end{equation}
where
\begin{equation}
\hat{Z}_{q}=\sum_{i=1}^{n}\Big(1+(q-1){\beta}(E_{i}-U_{q})/(\sum_{i=1}^{n}p_{i}^{q})
\Big)^{\frac{1}{1-q}},
\end{equation} continuously map  the set $D$ into itself.
\end{Pn}
Using Lefschetz fixed point theorem \cite{Lef}  we obtain the assertion:
\begin{Tm} 
Let relations \eqref{j1}  continuously map the set $D$ into itself. Then there exists one and only one point 
\begin{align}\nn &
\tilde{P}=(\tilde{p_{1}},\tilde{p_{2}},...,\tilde{p_{n}}),
\end{align}
which satisfies relations \eqref{4}, \eqref{5}, and $\tilde{P}{\in}D$.
\end{Tm}
\begin{proof}. It follows from the analyticity of $r_{i}$  that  the map under consideration has only a finite number $N_{f}$ of fixed points. Hence we can apply the Lefschetz fixed point theorem \cite{Lef}. According to this
theorem the number $N_{f}$  coincides with the Euler characteristics $\chi(D)$ of $D$. In view of the well-known Euler formula we have $\chi(D)=1$. The theorem is proved.
\end{proof}
\begin{Rk} Relations \eqref{j1} continuously map the set $D$ into itself
if either conditions of Proposition 2.1  or conditions of
Proposition 2.2 are fulfilled.
\end{Rk}
We stress, that  we consider  the extremal problem for the introduced function F, which contains the fixed parameter $\beta$, but the energy $U_{q}$ 
is not fixed. The case, where the energy $U_{q}$ is fixed, was treated in a number of works but the corresponding equation for the Lagrange multiplier is transcendental   and very complicated.

For the proof of our next proposition, we use the classical iteration method and 
take $P_{0}=(1/n,1/n,...,1/n)$ as the starting point.

\begin{Pn} If $\beta$ is small, then
\begin{equation}\tilde{p_{i}}{\approx}1/n+{\beta}(E_{i}-\bar{E})n^{-q},\end{equation}
where $\bar{E}=(\sum_{i=1}^{n}{E_{i}})/n.$
\end{Pn}

\section{Conclusion}
 The traditional approach to entropy (and the importance  of the rigorous  treatment of this notion, wherever possible)
was described by A. Wehrl \cite{Weh1} 
in the following statement:
``{\it Traditionally entropy is derived from phenomenological thermodynamical
considerations based upon the second law of thermodynamics. This method 
does not seem to be appropriate for a profound understanding of entropy}''.

In our note we address the problem of the rigorous treatment of the Gibbs-type
formulas closely related to entropy.
The conditional extremum problem that is treated in the note
could be considered as the game situation in accordance with
the definition that
"{\it game theory models strategic situations in which an individual's success in making choices depends  on the choices of
 others}" $($see, e.g., \cite{My}$)$.
 
 The rigorous proof  that under rather weak
conditions the probabilities given by the Gibbs-type
formula exist and are unique is both important and new.
 
The obtained results afford a rigorous treatment of other problems
of non-extensive mechanics and applications to other domains
(see, e.g., \cite{LAS0, LAS3}).

 \end{document}